# OTTICA GEOMETRICA CLASSICA: PROPAGAZIONE NEI MEZZI DISOMOGENEI.

(Metodo dell'iconale)

Silvio Bianchi[1] , Umberto Sciacca[2], Alessandro Settimi[2]

[1] Università Sapienza – Roma
[2] Istituto Nazionale di Geofisica e Vulcanologia – Roma

# INDICE

**Introduzione**

Una corretta caratterizzazione dei collegamenti radio su grandi distanze non può prescindere dallo studio degli effetti generati dalla presenza dell'atmosfera terrestre, in particolare della troposfera il cui indice di rifrazione diminuisce lentamente con la quota. In un radiocollegamento la propagazione avviene quindi in mezzi ad indice di rifrazione lentamente variabile. In generale si hanno i modi di propagazione di cui in fig. 1.

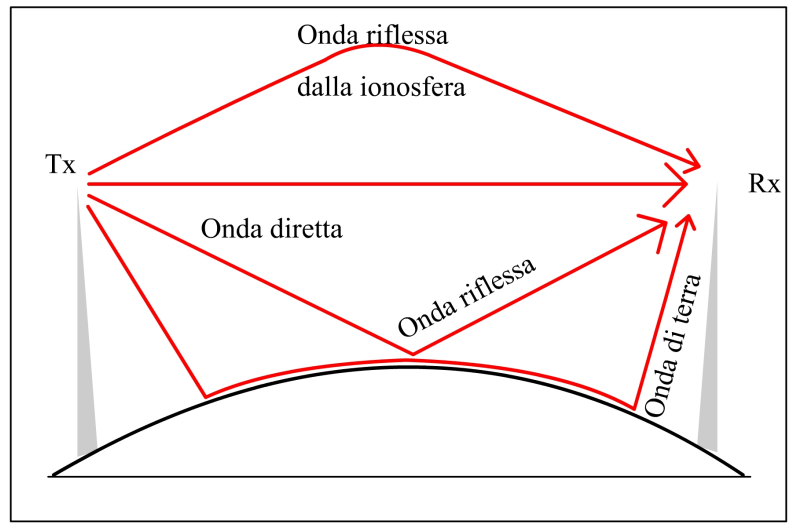

Figura 1 - Possibili modi di propagazione dell'onda elettromagnetica da un terminale Tx a un terminale Rx

Visto che l'andamento dell'indice di rifrazione varia, la traiettoria del segnale non è rettilinea. È necessario prevederla per valutare l'esatto puntamento delle antenne, l'esatta attenuazione, e l'esatto ritardo di propagazione. Una trattazione della propagazione in modo esatto a partire dalle equazioni di Maxwell è molto complessa ed è quindi necessario ricorrere ad una trattazione semplificata.

La natura ed il comportamento della luce ci consentono di interpretare alcuni fenomeni tramite i raggi luminosi, ognuno dei quali si può pensare come un segmento di retta che ha la direzione di propagazione del fronte d'onda. Tale modello, noto come "ottica geometrica", fu introdotto da Keplero, costituisce una buona approssimazione della realtà ed è di estrema utilità nello studio dei fenomeni di riflessione e rifrazione, nonché degli effetti prodotti dai vari tipi di specchi (piani, concavi e convessi) e dalle lenti.

I fenomeni interpretati mediante l'ottica geometrica possono essere spiegati anche con la teoria ondulatoria della luce sebbene con maggiore difficoltà nell'effettuare i calcoli. Il campo $f(\vec{r},t)$ in questo caso è soluzione descritto dalla cosiddetta equazione delle onde:

$$\nabla^2 f(\vec{r},t) - \frac{1}{v^2}\frac{\partial f(\vec{r},t)}{\partial t^2} = 0,$$

dove v è la velocità della luce nel mezzo considerato. La funzione d'onda può poi essere associata ad una delle componenti del campo elettromagnetico ed si possono applicare a tale componente tutte le considerazioni fatte nei corsi di Campi Elettromagnetici.

È importante sottolineare che questa è una teoria scalare e che quindi fa perdere la possibilità di descrivere tutti quei fenomeni, come ad esempio la polarizzazione, che richiedono la conoscenza di tutte le componenti del campo.

La teoria geometrica è un modello più semplice della teoria ondulatoria, è in grado di descrivere la propagazione in mezzi non omogenei inizialmente supposti senza perdite, a condizione che gli scostamenti dall'uniformità siano piccoli su lunghezze confrontabili con la lunghezza d'onda.

I casi in cui si manifestano i fenomeni tipici dell'ottica (riflessione, rifrazione e diffrazione), che sono accomunati dal fatto che il sistema interagente con il campo è costituito da oggetti di grandi dimensioni rispetto alla lunghezza d'onda, sono:

1. regioni caratterizzate da costanti costitutive del mezzo lentamente variabili nello spazio (tipicamente nell'atmosfera), come la permittività dielettrica $\varepsilon_r(\vec{r})$ o l'indice di rifrazione $n(\vec{r}) = \sqrt{\varepsilon_r(\vec{r})}$ (in questo caso l'ottica geometrica diviene sempre più precisa al crescere della frequenza);
2. estese superfici di materiale solido, come spigoli, pareti (in questo caso valgono le teorie delle riflessione, l'ottica fisica e la teoria geometrica delle diffrazione).

Le condizioni di lenta variabilità, ancorché necessarie, non sono sufficienti a garantire la totale esattezza delle soluzioni ottenute con l'ottica geometrica; possono infatti sussistere concentrazioni di raggi in regioni dette "fuochi" o "caustiche" in cui l'ottica geometrica porta a soluzioni incomplete o erronee.

L'insieme dei raggi rifratti, riflessi e diffratti costituisce una descrizione del campo sufficientemente approssimata in regioni complesse quali l'ambito urbano o extraurbano quando si voglia tener conto del profilo verticale dell'indice di rifrazione troposferica e/o delle irregolarità del terreno.

Rimangono escluse le onde superficiali, non configurabili come raggi, ma la cui importanza decresce all'aumentare della frequenza.

# 1. Richiami di ottica geometrica

## 1.1 Equazioni di base

Ricaveremo le equazioni fondamentali dell'ottica geometrica nell'ambito della trattazione scalare, partendo dall'equazione di Helmholtz scritta per un mezzo in cui l'indice di rifrazione possa variare da punto a punto. Indicando con $U(x,y,z,t)$ la funzione caratterizzante l'onda elettromagnetica, e con $n(x,y,z,t)$ l'indice di rifrazione, entrambe funzioni del punto di raggio vettore $\vec{r}=(x,y,z)$ e dell'istante $t$, si riscrive l'equazione delle onde:

$$\frac{\partial^2 U(x,y,z,t)}{\partial x^2}+\frac{\partial^2 U(x,y,z,t)}{\partial y^2}+\frac{\partial^2 U(x,y,z,t)}{\partial z^2}=\frac{n^2(x,y,z,t)}{c^2}\frac{\partial^2 U(x,y,z,t)}{\partial t^2}, \tag{1.1}$$

essendo $c$ la velocità della luce nel vuoto. Le proprietà spazio-temporali di un mezzo continuo sono caratterizzate da una funzione continua, l'indice di rifrazione $n(x,y,z,t)$; esso, nell'ipotesi che il mezzo sia invariante nel tempo, non dipenderà esplicitamente da $t$. Inoltre, se l'onda e.m. è monocromatica, cioè un'armonica pura esprimibile come $U(x,y,z)e^{-i\omega t}$, con pulsazione $\omega$, si può risolvere l'equazione (1.1) applicando il metodo della separazione nelle variabili spaziali $(x,y,z)$ e temporale $t$:

$$\frac{\partial^2 U(x,y,z)}{\partial x^2}+\frac{\partial^2 U(x,y,z)}{\partial y^2}+\frac{\partial^2 U(x,y,z)}{\partial z^2}+k_0^2 n^2(x,y,z)U(x,y,z)=0, \tag{1.2}$$

nota come equazione di Helmholtz, dove il vettore d'onda nel vuoto $k_0$ è definito tramite la pulsazione $k_0=\omega/c$ o la lunghezza d'onda $k_0=2\pi/\lambda$. Si osserva che l'equazione (1.2) vale nell'ipotesi di cambiamenti non troppo bruschi di $n(x,y,z)$, condizione che si verifica quando il mezzo è generalmente continuo.

Scriviamo $U(x,y,z)$ nella forma:

$$U(x,y,z)=A(x,y,z)e^{ik_0 S(x,y,z)}, \tag{1.3}$$

con $A(x,y,z)$ ed $S(x,y,z)$ reali. Come casi particolari, se l'indice di rifrazione $n$ è costante, quindi per mezzo omogeneo, tra le soluzioni esatte della (1.2) vi è quella che descrive l'onda piana:

$$U(\vec{r})=Ae^{ik_0 nr(\hat{n}\cdot\hat{r})},$$

dove $\hat{n}$ è il versore normale al fronte d'onda e $\hat{r}$ il versore parallelo alla direzione di propagazione. Altra soluzione (valida anche per mezzo isotropo) è l'onda sferica:

$$U(r)\propto\frac{1}{r}e^{ik_0 nr},$$

che si propaga indipendentemente dalla direzione.

Vediamo quale espressione assume la (1.2) inserendovi la (1.3). Differenziando $U(x,y,z)$ rispetto a $x$, si ottiene:

$$\frac{\partial U}{\partial x}=\frac{\partial A}{\partial x}\cdot e^{ik_0 S}+A\cdot e^{ik_0 S}ik_0\frac{\partial S}{\partial x}=[\frac{\partial A}{\partial x}+ik_0 A\frac{\partial S}{\partial x}]e^{ik_0 S},$$

dove per brevità si è omessa la dipendenza esplicita di $S$ e $A$ da $(x,y,z)$. Differenziando ancora $\partial U/\partial x$ rispetto ad $x$, si ottiene:

$$\frac{\partial^2 U}{\partial x^2} = [\frac{\partial^2 A}{\partial x^2} + ik_0(\frac{\partial A}{\partial x}\frac{\partial S}{\partial x} + A\frac{\partial^2 S}{\partial x^2})] \cdot e^{ik_0 S} + (\frac{\partial A}{\partial x} + ik_0 A\frac{\partial S}{\partial x}) \cdot e^{ik_0 S} ik_0 \frac{\partial S}{\partial x} =$$

$$= [\frac{\partial^2 A}{\partial x^2} + ik_0(\frac{\partial A}{\partial x}\frac{\partial S}{\partial x} + A\frac{\partial^2 S}{\partial x^2}) + ik_0 \frac{\partial S}{\partial x}(\frac{\partial A}{\partial x} + ik_0 A\frac{\partial S}{\partial x})]e^{ik_0 S} =$$

$$= [\frac{\partial^2 A}{\partial x^2} + 2ik_0 \frac{\partial A}{\partial x}\frac{\partial S}{\partial x} + ik_0 A\frac{\partial^2 S}{\partial x^2} - k_0^2 A(\frac{\partial S}{\partial x})^2]e^{ik_0 S}$$

Differenziando $U(x,y,z)$ anche rispetto a $y$ e $z$, si ottengono espressioni del tutto analoghe alla precedente, solo con le variabili $y$ e $z$ al posto di $x$, riportate nella seguente espressione, in cui si va a sommare membro a membro le tre derivate:

$$\frac{\partial^2 U}{\partial x^2} + \frac{\partial^2 U}{\partial y^2} + \frac{\partial^2 U}{\partial z^2} = \nabla^2 U =$$
$$= [\frac{\partial^2 A}{\partial x^2} + 2ik_0 \frac{\partial A}{\partial x}\frac{\partial S}{\partial x} + ik_0 A\frac{\partial^2 S}{\partial x^2} - k_0^2 A(\frac{\partial S}{\partial x})^2]e^{ik_0 S} +$$
$$+ [\frac{\partial^2 A}{\partial y^2} + 2ik_0 \frac{\partial A}{\partial y}\frac{\partial S}{\partial y} + ik_0 A\frac{\partial^2 S}{\partial y^2} - k_0^2 A(\frac{\partial S}{\partial y})^2]e^{ik_0 S} + \quad . \quad (1.4)$$
$$+ [\frac{\partial^2 A}{\partial z^2} + 2ik_0 \frac{\partial A}{\partial z}\frac{\partial S}{\partial z} + ik_0 A\frac{\partial^2 S}{\partial z^2} - k_0^2 A(\frac{\partial S}{\partial z})^2]e^{ik_0 S} =$$
$$= [\nabla^2 A + 2ik_0 \nabla A \cdot \nabla S + ik_0 A \nabla^2 S - k_0^2 A |\nabla S|^2]e^{ik_0 S}$$

Sostituendo le equazioni (1.4) e (1.3) nella (1.2), poiché l'esponenziale $e^{ik_0 S}$ non può essere uguale a zero, si deduce:

$$\nabla^2 A + 2ik_0 \nabla A \cdot \nabla S + ik_0 A \nabla^2 S - k_0^2 A |\nabla S|^2 + k_0^2 n^2 A =$$
$$= \nabla^2 A + k_0^2 A(n^2 - |\nabla S|^2) + ik_0(A \nabla^2 S + 2 \nabla A \cdot \nabla S) = 0$$

Separando la parte reale e la parte immaginaria, si ottengono le equazioni:

$$\nabla^2 A + k_0^2 A(n^2 - |\nabla S|^2) = 0 , \tag{1.5}$$

$$A \nabla^2 S + 2 \nabla A \cdot \nabla S = 0 . \tag{1.6}$$

Si noti che, con opportuna scelta di $A(\vec{r})$ ed $S(\vec{r})$, è sempre possibile porre $U(\vec{r})$ nella forma (1.3), perciò le (1.5) e (1.6) hanno validità generale.

A questo punto si può introdurre l'approssimazione che porta all'ottica geometrica; essa consiste nell'ammettere che nella (1.5) il termine $\nabla^2 A$ sia trascurabile rispetto agli altri due:

$$\nabla^2 A << k_0^2 A(n^2 - |\nabla S|^2) . \tag{1.7}$$

Poiché questi ultimi sono moltiplicati per $k_0^2$ e quindi hanno un peso che cresce come $\omega^2$ (ovvero $1/\lambda^2$), si usa dire che l'ottica geometrica è il limite di quella ondulatoria nell'ipotesi di frequenze relativamente alte $\omega \to \infty$ (ovvero di lunghezze d'onda basse $\lambda \to 0$).

Introducendo l'approssimazione detta (1.7) nella (1.5), poiché si esclude la soluzione banale in cui l'ampiezza $A$ è uguale a zero, si ottiene l'equazione:

$$\nabla S \cdot \nabla S = |\nabla S|^2 = n^2 . \tag{1.8}$$

La funzione $S(\vec{r})$ si chiama iconale, mentre l'equazione (1.8) è nota come *equazione dell'iconale*; essa descrive la variazione spaziale della fase e la mette in relazione con l'indice di rifrazione. Risolta tale equazione, ci si può servire della (1.6), nota come equazione di trasporto, per trovare la funzione di ampiezza $A(\vec{r})$.

Ci interessiamo per ora della equazione (1.8). Le superfici $S(\vec{r}) = \text{cost}$ sono le *superfici equifase* dell'onda e.m $U(\vec{r})$ [vedi eq. (1.3)], mentre le "linee di forza" $\nabla S$ vengono prese per definizione come *raggi luminosi*. In altre parole, se $S(x,y,z)=cost$ è una superficie equifase, allora le derivate parziali $\partial S/\partial x$, $\partial S/\partial y$, $\partial S/\partial z$ sono le componenti di un vettore normale alla superficie equifase, cioè il gradiente dell'iconale con modulo pari all'indice di rifrazione $n(x,y,z)$. Il collegamento tra questa definizione di raggio luminoso ed il corrispondente concetto intuitivo sarà chiarito nel seguito. Da quanto abbiamo detto risulta dunque che i raggi luminosi sono le traiettorie ortogonali alle superfici d'onda. Una famiglia di curve di questo tipo viene detta *congruenza normale*, in particolare è detta congruenza rettilinea quando i raggi sono rette.

Volendo prevedere anche l'andamento dell'ampiezza dell'onda, consideriamo un generico raggio luminoso ed introduciamo su di esso un'ascissa. Detto $\hat{s}$ il versore della tangente al raggio, dato che $\hat{s} = \nabla S / n = d\vec{r}/ds$, l'equazione dell'iconale può scriversi:

$$\nabla S = n\hat{s} = n\, d\vec{r}/ds . \tag{1.9}$$

Se riscriviamo la (1.6), dividendola per $A$, essa si può scrivere nella forma:

$$\nabla^2 S + \nabla S \cdot \nabla(\ln A^2) = 0 ;$$

inserendovi l'equazione dell'iconale (1.9), si ottiene:

$$\nabla^2 S + n\hat{s} \cdot \nabla(\ln A^2) = \nabla^2 S + n\frac{d}{ds}(\ln A^2) = 0 ,$$

(in questo contesto il termine $\hat{s} \cdot \nabla[\ ]$ assume il significato di derivata direzionale $d[\ ]/ds$ dato che il versore $\hat{s}$ è tangente all'ascissa curvilinea *s*), per cui:

$$\frac{d}{ds}(\ln A^2) = -\frac{\nabla^2 S}{n} . \tag{1.10}$$

L'integrazione numerica dell'equazione differenziale (1.10) porta alla determinazione dell'ampiezza d'onda $A(\vec{r})$ lungo il raggio luminoso *s*. La (1.10), insieme alla (1.9), costituisce un'approssimazione migliore rispetto alla sola equazione dell'iconale (1.8) in quanto tiene conto anche delle variazioni dell'ampiezza $A(\vec{r})$.

**1.2. Interpretazione grafica dell'equazione dell'iconale**

L'equazione dell'iconale (1.9) è passibile di un'interpretazione grafica, per la quale ci si avvale dell'ausilio della fig. 2. Vale l'ipotesi di propagazione in un mezzo non omogeneo, cosa che comporta un percorso del raggio curvo (riportato in fig. 2, ove è segnata la coordinata curvilinea *s*).

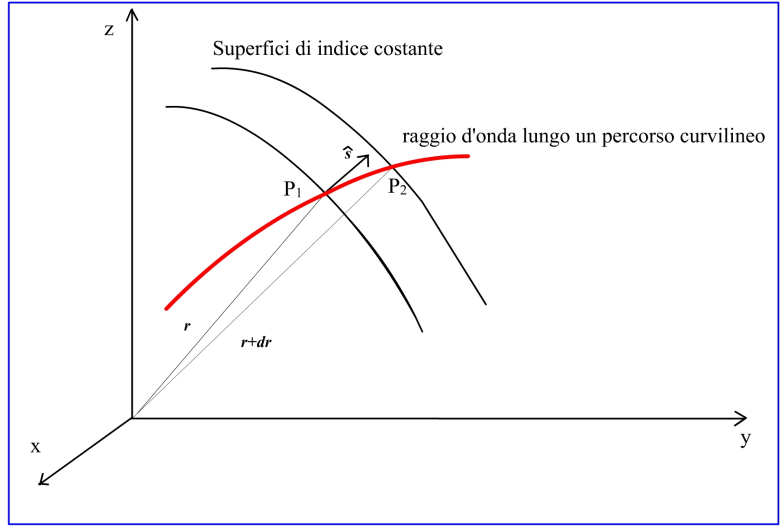

Figura 2 - Il percorso del raggio luminoso nel punto P intercetta la superficie equifase dell'indice di rifrazione nel punto P. Il versore $\hat{s}$ è tangente al raggio d'onda e normale alla superficie equifase.

Con riferimento all'equazione dell'iconale (1.9), il campo vettoriale $n\hat{s}$ è dunque il gradiente della funzione scalare iconale $S$ e quindi è conservativo ($\nabla \times (\nabla f) = 0$ per qualunque funzione *f*). Lungo qualunque linea chiusa *l* si ha allora:

$$\oint n\hat{s} \cdot \vec{dl} = 0, \quad (1.11)$$

essendo $\vec{dl}$ dl il generico elemento della linea *l*. L'integrale curvilineo (1.11) è noto come *invariante integrale di Lagrange*. Presi due punti $P_1$ e $P_2$ con raggi vettori $\vec{r}_1$ ed $\vec{r}_2$ e scelta una qualunque linea che li congiunga, in base alla eq. (1.9) si ha:

$$\int_{P_1}^{P_2} n\hat{s} \cdot \vec{dl} = \int_{P_1}^{P_2} \nabla S \cdot \vec{dl} = S_2 - S_1, \quad (1.12)$$

avendo posto $S(\vec{r}_1) = S_1$ e $S(\vec{r}_2) = S_2$. Supponiamo ora che $P_1$ e $P_2$ si trovino su uno stesso raggio luminoso (fig. 2). In questo caso è $n\hat{s} \cdot \vec{dl} = ds$ e la (1.12) diventa:

$$\int_{P_1}^{P_2} n\, ds = S_2 - S_1. \quad (1.13)$$

Ammettiamo che un solo raggio luminoso passi per $P_1$ e $P_2$; presa allora una qualunque altra linea *l* che colleghi $P_1$ e $P_2$, lungo tale linea sarà $\hat{s} \cdot \vec{dl} < dl$ e quindi [vedi eq. (1.12)]:

$$S_2 - S_1 = \int_{P_1}^{P_2} n\hat{s} \cdot \vec{dl} \leq \int_{P_1}^{P_2} n\, dl. \quad (1.14)$$

La quantità

$$\int_{P_1}^{P_2} n\, dl \quad (1.15)$$

lungo una generica linea, o in particolare lungo un raggio luminoso, prende il nome di cammino ottico lungo la linea detta. Nel caso in cui la linea coincida con un raggio luminoso, il cammino ottico è uguale alla variazione di iconale fra i due punti estremi considerati [vedi eq. (1.13)], mentre per un'altra linea esso è maggiore di tale variazione.

L'equazione (1.14) giustifica il celebre principio di Fermat, enunciato nel 1657 (quindi ben prima della teoria elettromagnetica della luce) e noto come principio del minimo cammino ottico: *"Il cammino ottico lungo il raggio luminoso che congiunge due punti è minimo rispetto a tutti i percorsi che congiungono i due punti"* (in realtà, l'enunciato di Fermat era più teleologico: *"La natura segue sempre le vie più brevi"*). Tenendo presente la definizione dell'indice di rifrazione, si vede che la quantità *ndl* è proporzionale al tempo impiegato dalla luce per compiere il tratto *dl*; per questa ragione, il principio di Fermat va anche sotto il nome di principio del minimo tempo. Tuttavia in altri casi si può trovare addirittura che il percorso seguito dal raggio corrisponde ad un massimo locale del cammino ottico. Una formulazione del principio valida in generale asserisce che il cammino ottico lungo un raggio è stazionario nel senso del calcolo delle variazioni, cioè che la cosiddetta variazione prima dell'integrale (1.15) è nulla per piccole deformazioni del percorso di integrazione rispetto a quel raggio.

Per quanto riguarda la rifrazione, è facile stabilire la legge di Snell-Cartesio applicando il principio di Fermat con il seguente procedimento (vedi fig. 3).

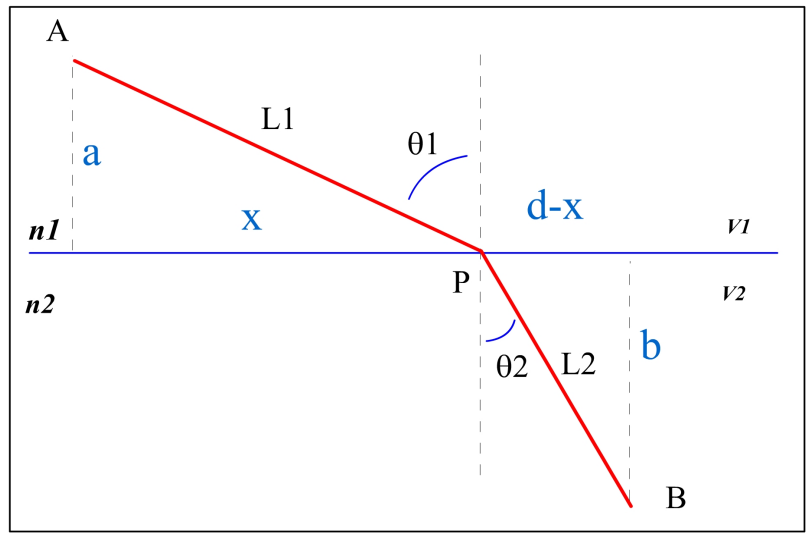

Figura 3 – Semplice schema che mostra la minimizzazione del tempo di propagazione.

Dati due punti *A* e *B* in due mezzi omogenei con indici di rifrazione $n_1$ e $n_2$ separati da una superficie, che per il momento supponiamo piana, consideriamo il piano passante per i due punti e ortogonale alla superficie di separazione (tale piano costituisce il piano del disegno nella figura). Il cammino ottico da *A* a *B* lungo un percorso formato dai due segmenti *AP* e *PB* è funzione della posizione di *P* sulla superficie di separazione. Con le notazioni di figura 3, tale cammino, diciamo $L_{AB}(x)$, può esprimersi come:

$$L_{AB}(x) = n_1 \sqrt{a^2 + x^2} + n_2 \sqrt{b^2 + (d-x)^2} \,, \tag{1.16}$$

dove *d* è la distanza fra i piedi delle perpendicolari abbassate da *A* e *B* sulla superficie di separazione.

Se il percorso considerato coincide con quello del raggio, il cammino corrispondente deve essere stazionario. Nel caso in esame ciò significa semplicemente che si deve annullare la derivata rispetto ad *x*. Imponendo questa condizione, si ritrova subito la legge di Snell-Cartesio:

$$n_1 \sin \varphi_1 = n_2 \sin \varphi_2 \,. \tag{1.17}$$

Ci si rende conto inoltre che l'ipotesi di superficie di separazione piana è inessenziale e che la legge vale per una superficie di separazione di forma qualsiasi (almeno nei punti in cui la superficie è dotata di piano tangente).

**1.3. Andamento dell'intensità luminosa – Tubi di flusso**

La funzione d'onda $U$ può essere pensata come rappresentativa del campo elettrico dell'onda e.m.. Ricordiamo che per un'onda piana sinusoidale l'intensità luminosa media ha l'espressione (potenza istantanea dell'onda):

$$I = \frac{E^2}{2}\sqrt{\frac{\varepsilon_0 \varepsilon_r}{\mu_0 \mu_r}} \cong \frac{nE^2}{2Z_0}, \qquad (1.18)$$

essendo $E$ l'ampiezza del campo elettrico, $\varepsilon_0$ e $\mu_0$ la costante dielettrica e la permeabilità magnetica del vuoto, $\varepsilon_r$ e $\mu_r$ le analoghe permittività relative del mezzo. Nell'ultimo passaggio si è supposto $\mu_r \cong 1$ (come è ragionevole per mezzi otticamente trasparenti) e si è indicata con $Z_0 = \sqrt{\mu_0/\varepsilon_0}$ la cosiddetta impedenza caratteristica del vuoto. In analogia con la eq. (1.18) e trascurando le costanti di proporzionalità, poniamo:

$$I = nA^2. \qquad (1.19)$$

L'andamento dell'intensità luminosa in ottica geometrica può essere studiato a partire dall'equazione di trasporto (1.6) che regola il comportamento dell'ampiezza $A$. Inserendovi l'equazione (1.19), riscriviamo l'equazione di trasporto (1.10) nella forma:

$$\frac{d}{ds}(\ln\frac{I}{n}) = -\frac{\nabla^2 S}{n}. \qquad (1.20)$$

Integrando l'equazione (1.20) fra due punti generici di ascisse curvilinee $s_1$ e $s_2$, otteniamo:

$$I(s_2) = \frac{n(s_2)}{n(s_1)} I(s_1) e^{-\int_{s_1}^{s_2}\frac{\nabla^2 S}{n}ds}. \qquad (1.21)$$

L'equazione (1.21) permette il calcolo dell'intensità lungo un generico raggio se è stata risolta l'equazione dell'iconale (1.9), in modo che $\nabla^2 S$ sia una funzione nota. Osservando la (1.21), si nota che se l'intensità si annulla in un punto di un raggio, allora si annulla lungo tutto il raggio; ciò suggerisce l'idea che l'energia luminosa si propaghi lungo i raggi.
Se si moltiplica l'equazione (1.6) per $A$, si può scrivere:

$$A^2 \nabla^2 S + 2A\nabla A \cdot \nabla S = \nabla \cdot (A^2 \nabla S) = 0, \qquad (1.22)$$

che, utilizzando le (1.9) e (1.19), diventa:

$$\nabla \cdot (I\hat{s}) = 0. \qquad (1.23)$$

Dunque il campo $I\hat{s}$ è solenoidale. Ora, si può ricavare la legge di intensità dell'ottica geometrica, che è un'evoluzione dell'espressione dell'intensità in termini di tubi di flusso.
Si definisce tubo di flusso dell'energia una superficie chiusa costituita lateralmente da una famiglia di raggi ed ortogonalmente da due porzioni di superficie d'onda (vedi fig. 4).

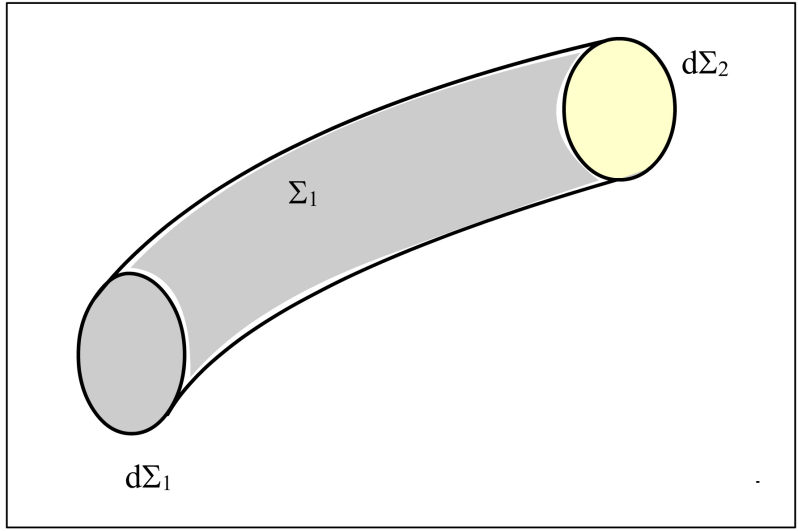

Figura 4 – Tubo di flusso

Nell'ipotesi di mezzo privo di perdite, applichiamo il teorema di Poynting ad un tubo di flusso di sezioni sufficientemente piccole $d\Sigma_1$ e $d\Sigma_2$, tali da poter considerare su di esse il vettore di Poynting quasi costante ($\vec{\Pi} \cong \text{cost.}$); per la conservazione dell'energia si ha:

$$\oint_\Sigma \vec{\Pi} \cdot \hat{n} d\Sigma = \int_{d\Sigma_1} \vec{\Pi} \cdot \hat{n} d\Sigma + \int_{d\Sigma_2} \vec{\Pi} \cdot \hat{n} d\Sigma + \underbrace{\int_{\Sigma_l} \vec{\Pi} \cdot \hat{n} d\Sigma}_{=0} = -\left|\vec{\Pi}_1\right| d\Sigma_1 + \left|\vec{\Pi}_2\right| d\Sigma_2 = 0; \quad (1.24)$$

da cui si deduce la legge di intensità dell'ottica geometrica:

$$I_1 d\Sigma_1 = I_2 d\Sigma_2. \quad (1.25)$$

L'energia luminosa viaggia all'interno del tubo di flusso e l'intensità varia in modo inversamente proporzionale alla sezione del tubo. Si definisce il fattore di divergenza *u* come:

$$u^2 = \frac{\left|\vec{\Pi}_2\right|}{\left|\vec{\Pi}_1\right|} = \frac{I_2}{I_1} = \frac{d\Sigma_1}{d\Sigma_2}. \quad (1.26)$$

Tale grandezza tiene conto dell'eventuale attenuazione dovuta all'allargamento del fronte d'onda con la propagazione. L'intensità portata da ogni raggio può diminuire con la distanza anche se il mezzo é privo di perdite poiché, man mano che l'onda avanza, l'energia viene distribuita su una superficie sempre più ampia.

Si faccia ora riferimento alla figura 5, ove $C_1C_2$ e $C_3C_4$ sono le cosiddette *caustiche* dell'onda. In generale, la superficie costituita dall'insieme dei punti in cui i raggi convergono si chiama caustica e nel caso tale superficie si riduca ad un punto, i raggi convergono su di esso, che viene detto fuoco.

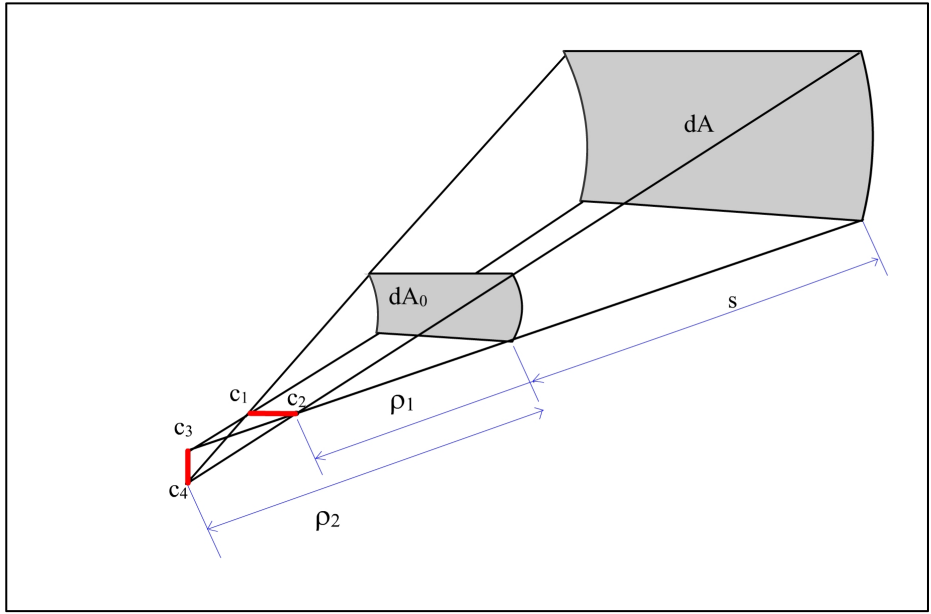

Figura 5 – Caustiche dell'onda

Nell'ipotesi di mezzo omogeneo, quando un'onda generica si propaga per raggi rettilinei, si può dimostrare che il fattore di divergenza assume l'espressione ($\rho_1$ e $\rho_2$ sono i raggi di curvatura principali dell'onda):

$$u(\rho_1, \rho_2, s) = \sqrt{\frac{\rho_1 \rho_2}{(\rho_1 + s)(\rho_2 + s)}} . \tag{1.27}$$

Si possono distinguere i tre casi usuali di riferimento:
1. onda sferica

$$\rho_1 = \rho_2 = \rho_0 \quad \Rightarrow \quad u = \frac{\rho_0}{\rho_0 + s} ; \tag{1.28}$$

2. onda cilindrica

$$\rho_1 = \infty \quad , \quad \rho_2 = \rho_0 \quad \Rightarrow \quad u = \sqrt{\frac{\rho_0}{\rho_0 + s}} ; \tag{1.29}$$

3. onda piana

$$\rho_1 = \rho_2 = \infty \quad \Rightarrow \quad u = 1 . \tag{1.30}$$

Il fattore di divergenza permette di capire come evolve l'ampiezza e quindi l'intensità del campo lungo il raggio ottico. Infatti, se il mezzo è omogeneo, si può risolvere l'equazione differenziale (1.20), ottenendo che la propagazione dell'onda e.m. lungo un raggio è descritta dalla seguente relazione:

$$U(s) = \underbrace{U(0)}_{\text{onda e.m. nel punto } s=0} \cdot \underbrace{u(\rho_1, \rho_2, s)}_{\text{fattore di divergenza}} \cdot \underbrace{e^{ikS}}_{\text{fattore di fase}} . \tag{1.31}$$

La (1.31) permette quindi di calcolare il campo in un punto $P_2$, partendo dal valore del campo nel punto $P_1$, da cui esce il raggio passante per $P_2$. Il suo uso non è ammissibile in prossimità dei punti in cui la sezione del fascio infinitesimo preso intorno al raggio considerato si annulla, cioè per un fascio omocentrico (tutti i raggi passano per un punto); in questo caso infatti il fattore di divergenza tende all'infinito e il campo calcolato diverge. Questo risultato, privo di senso fisico, è dovuto al fatto che in prossimità del centro del tubo di flusso, dove l'ampiezza A varia molto rapidamente, non è più lecita l'approssimazione (1.7) di trascurare $\nabla^2 A$, fatta nel passaggio dalla equazione (1.5) alla (1.8). In tali regioni, quindi in prossimità delle caustiche e dei fuochi, i risultati dell'ottica geometrica sono inaccettabili, perciò si dovrà usare l'ottica ondulatoria.

## 2. Equazione dei raggi

### 2.1. Caso generale

Riscriviamo l'eq. dell'iconale (1.8):

$$\nabla S \cdot \nabla S = |\nabla S|^2 = n^2,$$

ossia l'equazione (1.9): $\nabla S = n\hat{s}$, dove $n(\vec{r})$ è l'indice di rifrazione, $S(\vec{r})$ è l'iconale, e $\hat{s}$ è il versore tangente al raggio luminoso con ascissa curvilinea $s$, normale alla superficie equifase $S(\vec{r}) = $ cost .

Derivando la (1.9) rispetto a $s$, otteniamo

$$\frac{d}{ds}(n\hat{s}) = \frac{d}{ds}\nabla S = \frac{d}{ds}(\frac{\partial S}{\partial x}\hat{x} + \frac{\partial S}{\partial y}\hat{y} + \frac{\partial S}{\partial x}\hat{z}), \tag{2.1}$$

dove $\hat{x}$, $\hat{y}$, $\hat{z}$ sono i versori che generano lo spazio cartesiano.

Per ora, si prende in considerazione solo la componente lungo la direzione $x$, per cui si opera sul primo termine del secondo membro della (2.1).

Assegnata una generica funzione $f(x,y,z)$, il suo differenziale totale è definito come $df = (\partial f/\partial x)dx + (\partial f/\partial y)dy + (\partial f/\partial z)dz = \nabla f \cdot d\vec{r}$. Anche per la derivata parziale di $S(x,y,z)$ rispetto ad $x$, cioè $\partial S/\partial x$, si può calcolare il suo differenziale che è $d(\partial S/\partial x) = \nabla(\partial S/\partial x) \cdot d\vec{r}$; ne segue:

$$\frac{d}{ds}(\frac{\partial S}{\partial x}) = \nabla(\frac{\partial S}{\partial x}) \cdot \frac{d\vec{r}}{ds}.$$

Se $\hat{s}$ è il versore della tangente al raggio luminoso, allora per definizione $\hat{s} = d\vec{r}/ds$; segue che, applicando la (1.9), si può riscrivere la precedente come:

$$\frac{d}{ds}(\frac{\partial S}{\partial x}) = \nabla(\frac{\partial S}{\partial x}) \cdot \hat{s} = \nabla(\frac{\partial S}{\partial x}) \cdot \frac{\nabla S}{n}.$$

Applicando ancora la (1.9) e svolgendo dei passaggi intermedi:

$$\frac{d}{ds}(\frac{\partial S}{\partial x}) \stackrel{[eq.(1.9)]}{=} \frac{d}{ds}(n\frac{dx}{ds}) \doteq \nabla(\frac{\partial S}{\partial x}) \cdot \frac{\nabla S}{n} = \frac{1}{2n}\frac{\partial}{\partial x}(\nabla S \cdot \nabla S) \stackrel{[eq.(1.9)]}{=} \frac{1}{2n}\frac{\partial n^2}{\partial x} = \frac{\partial n}{\partial x} \quad ,$$

si trova la proiezione dell'equazione dei raggi lungo la direzione $x$:

$$\frac{d}{ds}(\frac{\partial S}{\partial x}) = \frac{d}{ds}(n\frac{dx}{ds}) = \frac{\partial n}{\partial x} . \tag{2.2}$$

Si procede analogamente per trovare la proiezione dell'equazione dei raggi lungo le direzioni $y$ e $z$:

$$\frac{d}{ds}(\frac{\partial S}{\partial y}) = \frac{d}{ds}(n\frac{dy}{ds}) = \frac{\partial n}{\partial y}, \tag{2.3}$$

$$\frac{d}{ds}(\frac{\partial S}{\partial z}) = \frac{d}{ds}(n\frac{dz}{ds}) = \frac{\partial n}{\partial z} \ . \tag{2.4}$$

Inserendo le ultime tre espressioni nella (2.1), si ricava l'equazione dei raggi, in una forma indipendente dal sistema di riferimento cartesiano:

$$\frac{d}{ds}(n\hat{s}) = \frac{d}{ds}(\nabla S) = \frac{d}{ds}(n\frac{d\vec{r}}{ds}) = \nabla n \ . \tag{2.5}$$

L'equazione differenziale (2.5) è molto importante in ottica geometrica classica poiché tiene conto della variazione del versore del raggio luminoso $\hat{s}$ al variare dell'indice di rifrazione $n(\vec{r})$. La (2.5) può essere integrata numericamente tramite il metodo di Runge-Kutta; caso per caso, si dovrà poi individuare la congruenza di raggi che soddisfa le condizioni iniziali del particolare problema in esame. Ad esempio, per un mezzo omogeneo, in cui si ha una congruenza rettilinea, a seconda del problema la soluzione può essere un fascio di raggi paralleli, un insieme di rette divergenti da un punto o una famiglia di rette senza un punto in comune né al finito né all'infinito, come accade per i raggi uscenti da uno strumento che non formi immagini stigmatiche.

Generalmente, se il mezzo non è omogeneo, allora il raggio luminoso procede su linee curve. In particolare, se $\nabla n = 0$, cioè l'indice di rifrazione $n$ è costante, ed il mezzo è omogeneo, allora $d(n\hat{s})/ds = 0$, cioè il versore $\hat{s}$ mantiene direzione e verso costanti, quindi il raggio procede su una linea retta.

Consideriamo ora il caso di un mezzo con indice di rifrazione a simmetria sferica, cioè in cui l'indice di rifrazione dipenda solo dalla distanza $r$ da un punto $O$ (centro di simmetria del mezzo):

$$n = n(r) \ , \tag{2.6}$$

per cui il $\nabla n$ avrà sempre la direzione di $r$. Prendiamo $O$ come origine dei vettori di posizione $\vec{r}$ e deriviamo rispetto ad $s$ la quantità $\vec{r} \times n\hat{s}$:

$$\frac{d}{ds}(\vec{r} \times n\hat{s}) = \frac{d\vec{r}}{ds} \times n\hat{s} + \vec{r} \times \frac{d}{ds}(n\hat{s}) \ . \tag{2.7}$$

Essendo $d\vec{r}/ds = \hat{s}$, il primo termine a secondo membro è nullo, mentre il secondo termine può scriversi, in base alla (2.5):

$$\vec{r} \times \frac{d}{ds}(n\hat{s}) = \vec{r} \times \nabla n \ , \tag{2.8}$$

quindi risulta anch'esso nullo in base all'ipotesi (2.6). Pertanto si ha:

$$\vec{r} \times n\hat{s} = \vec{c} \ , \tag{2.9}$$

dove $\vec{c}$ è un vettore costante. Secondo la (2.9) i raggi luminosi sono curve giacenti in piani che passano per il centro di simmetria. Inoltre, detto $\varphi$ l'angolo tra $\vec{r}$ e $\hat{s}$ nel generico punto di un raggio, si ha, lungo tutto il raggio:

$$nr \sin \varphi = \text{cost} \ . \tag{2.10}$$

Le equazioni (2.9) e (2.10) vanno sotto il nome di legge di Bouguer. Tale legge è analoga a quella della conservazione del momento della quantità di moto per una particella in moto sotto l'azione di una forza centrale. Infatti, tale relazione (legge di Snell per mezzi a simmetria sferica) è alla base della propagazione ionosferica o per onda di cielo, che sfrutta cioè la possibilità di avere rientro a terra oltre l'orizzonte geometrico di onde lanciate con elevazione maggiore di zero.

**2.2. Mezzo stratificato**

Come caso limite di mezzi con indice di rifrazione a simmetria sferica si può considerare quello di mezzi con indice di rifrazione dipendente da una sola variabile cartesiana, diciamo *z* (è il caso in cui il centro di simmetria si allontana indefinitivamente).

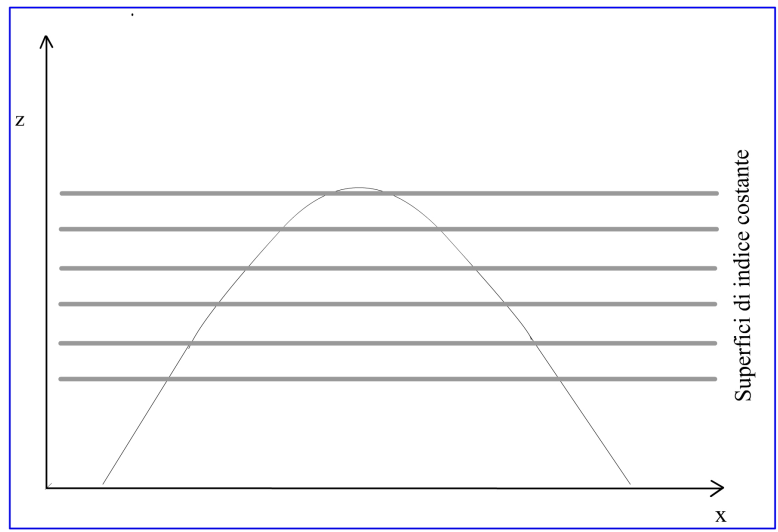

Figura 6 - Mezzo a stratificazione piana

Fissiamo un asse *x*, ortogonale a *z*, nel piano di un generico raggio che passi per un dato punto *($x_0$, $z_0$)* formando un angolo $\varphi_0$. Lungo il percorso del raggio, l'angolo $\varphi$ che esso forma con l'asse *z* cambierà. Tale angolo dipende quindi da *x*, *z*, e dalle condizioni iniziali. Per la legge di Bouguer avremo (fig.6):

$$n(z)\sin\varphi = n(z_0)\sin\varphi_0 = C ; \qquad (2.11)$$

essa generalizza la legge di Snell-Cartesio e può pensarsi ottenuta da questa nel caso di una successione di strati piani paralleli di spessore infinitesimo.

Dalle (2.11) si può ricavare l'equazione *z=z(x)* della traiettoria seguita da un generico raggio. Notiamo infatti che è:

$$\frac{dz}{dx} = \frac{1}{\tan\varphi} \qquad (2.12)$$

e che, limitandoci a considerare raggi con angolo $\varphi$ compreso fra zero e $\pi$, possiamo ricavare dalla (2.11):

$$\cos\varphi = \sqrt{1 - \frac{C^2}{n^2(z)}} . \qquad (2.13)$$

Inserendo le equazioni (2.11) e (2.13) nella (2.12) otteniamo l'equazione differenziale:

$$\frac{dz}{dx} = \frac{1}{C}\sqrt{n^2(z) - C^2} . \qquad (2.14)$$

Essa, una volta assegnata la funzione *n(z)*, si risolve per separazione di variabili nel seguente modo:

$$x = \int \frac{dz}{\sqrt{[\frac{n(z)}{C}]^2 - 1}}, \qquad (2.15)$$

dove l'integrale è indefinito e quindi contiene una costante di integrazione da determinare imponendo le condizioni iniziali, cioè che il raggio passi per il punto assegnato *(x₀, z₀)* formando con *z* l'angolo assegnato *φ₀*.

### 2.3. Derivazione della legge di Snell in termini differenziali

L'equazione dei raggi (2.5) può essere usata per derivare la legge di Snell-Cartesio in termini differenziali. Essa si enuncia nel modo seguente: in un mezzo disomogeneo, quando il raggio luminoso compie un percorso infinitesimo da un punto $\vec{r}$ con indice di rifrazione $n(\vec{r})$ a un punto $\vec{r}+d\vec{r}$ con indice di rifrazione $n(\vec{r}+d\vec{r}) = n(\vec{r}) + dn(\vec{r})$ (fig.7), il raggio viene rifratto in accordo con la seguente legge:

$$n(\vec{r}) \cdot \sin\varphi(\vec{r}) = [n(\vec{r}) + dn(\vec{r})] \cdot \sin[\varphi(\vec{r}) + d\varphi(\vec{r})], \qquad (2.16)$$

dove $\varphi(\vec{r})$ e $\varphi(\vec{r}) + d\varphi(\vec{r})$ sono gli angoli di rifrazione, rispettivamente definiti nei punti $\vec{r}$ e $\vec{r}+d\vec{r}$.

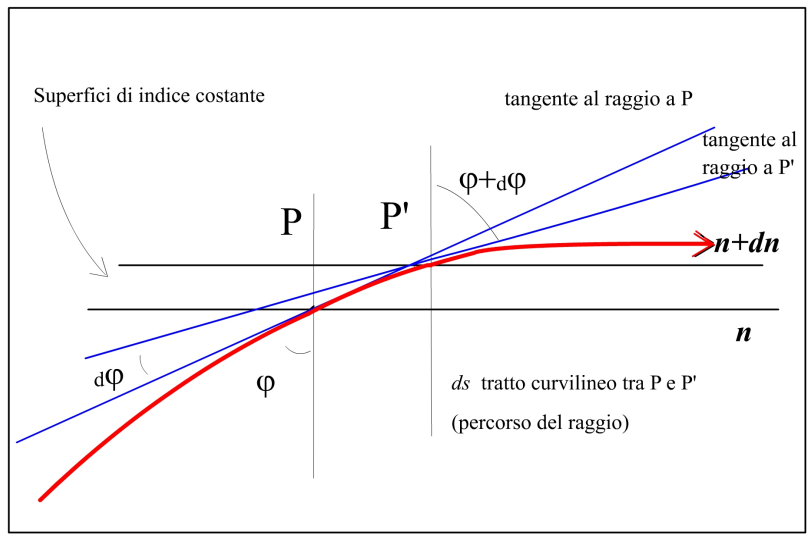

Figura 7 – Percorso del raggio luminoso tra i punti $\vec{r}$ e $\vec{r}+d\vec{r}$.

In fig. 7 i versori $\hat{s}_\varphi$ e $\hat{s}_{\varphi+d\varphi}$ formano rispettivamente gli angoli *φ* e *φ+dφ* con il versore $\hat{s}$ normale alla superficie equifase dell'indice di rifrazione $S(\vec{r}) = $ cost (In figura le superfici equifase dell'indice di rifrazione sono state disegnate come dei piani, ma in generale si possono immaginare come superfici curve).

In bibliografia esiste una dimostrazione rigorosa per l'enunciato appena esposto. qui se ne riportano sommariamente le linee guida. In accordo con l'equazione dei raggi (2.5), il versore $\hat{s}$ tangente al raggio luminoso risulta normale alla superficie equifase dell'indice di rifrazione, nel punto $\vec{r}$ dove raggio e superficie equifase si intercettano.

Con riferimento alla figura 7, se si risolve la (2.5) lungo un percorso infinitesimo tra i punti $\vec{r}$ e $\vec{r}+d\vec{r}$:

$$[n(\vec{r}) + dn(\vec{r})]\hat{s}_{\varphi+d\varphi} - n(\vec{r})\hat{s}_\varphi = |\nabla n|\hat{s}ds, \qquad (2.17)$$

dove $\hat{s}$ è il versore normale alla superficie equifase dell'indice di rifrazione $S(\vec{r}) = $ cost.

Si definisce il versore $\hat{\tau}$ tangente alla superficie equifase dell'indice di rifrazione che risulta normale al raggio luminoso, nel punto $\vec{r}$ dove superficie equifase e raggio si intercettano. Il versore $\hat{\tau}$ è normale al versore $\hat{s}$ che forma gli angoli $\varphi(\vec{r})$ e $\varphi(\vec{r})+d\varphi(\vec{r})$ rispettivamente con i versori $\hat{s}_{\varphi}$ e $\hat{s}_{\varphi+d\varphi}$. Se si proietta la (2.17) lungo la direzione del versore $\hat{\tau}$, dato che il contributo del secondo membro è nullo:

$$[n(\vec{r})+dn(\vec{r})]\hat{s}_{\varphi+d\varphi}\cdot\hat{\tau}-n(\vec{r})\hat{s}_{\varphi}\cdot\hat{\tau}=$$
$$=[n(\vec{r})+dn(\vec{r})]\sin[\varphi(\vec{r})+d\varphi(\vec{r})]-n(\vec{r})\sin\varphi(\vec{r})=, \qquad (2.18)$$
$$=|\nabla n|\hat{s}ds\cdot\hat{\tau}=0$$

si ottiene la (2.16).

**2.4. Esempio**

Come esempio di applicazione di quanto detto, studiamo il fenomeno del miraggio, ben noto perché si osserva frequentemente, per esempio sulle strade d'estate. Le zone dell'asfalto più lontane dall'osservatore sembrano bagnate, salvo risultare perfettamente asciutte quando l'osservatore si avvicina ad esse. In realtà tali zone si comportano come riflettenti; l'attribuire questa riflessione alla presenza di acqua è un'interpretazione legata a due fattori: a) la nostra radicata ma errata convinzione che i raggi luminosi debbano per forza procedere in linea retta e che possano deviare solo se riflessi; b) l'identificazione fra quello che osserviamo e ciò che abbiamo altre volte osservato in presenza di chiazze d'acqua sul terreno.

Fisicamente, ciò che accade è che gli strati d'aria prossimi all'asfalto sono più caldi e quindi meno densi di quelli sovrastanti, di conseguenza si ha un gradiente verticale dell'indice di rifrazione. I raggi luminosi si incurvano, cosicché un raggio, inizialmente diretto verso il basso, può arrivare a incurvarsi verso l'alto prima di incontrare il terreno, raggiungendo l'osservatore come se fosse stato riflesso. Dato che i gradienti in gioco sono piccoli, solo i raggi con un angolo $\varphi_0$ prossimo a $\pi/2$ subiscono il fenomeno; è per questo che, avvicinandosi, si ha la sensazione che l'acqua sparisca.

Per trovare la forma dei raggi adottiamo un modello semplice per $n(z)$:

$$n(z)=n_{min}+\beta z, \qquad (2.19)$$

in cui $n_{min}$ è il valore minimo dell'indice di rifrazione e $\beta$ una costante opportuna. Naturalmente penseremo che la (2.19) valga in un certo intervallo $0\leq z \leq z_{max}$. Introduciamo la (2.19) nella (2.15), operando il seguente cambiamento di variabile:

$$Z=\frac{n_{min}+\beta z}{C}. \qquad (2.20)$$

La (2.15) diventa allora:

$$\frac{\beta x}{C}=\int\frac{dZ}{\sqrt{Z^2-1}},$$

$$\frac{\beta(x-x_0)}{C}=\int_{z_0}^{z}\frac{dZ}{\sqrt{C^2Z^2-1}} \qquad (2.21)$$

la cui soluzione è:

$$\ln(Z+\sqrt{Z^2-1})=\frac{\beta x}{C}+\text{cost}, \qquad (2.22)$$

ovvero:

$$Z+\sqrt{Z^2-1} = Ae^{\frac{\beta x}{C}}, \qquad (2.23)$$

dove la costante *A* è da determinare per ogni particolare raggio conoscendo le coordinate *(x₀, z₀)* di un suo punto *P* e il corrispondente valore di *φ₀*. Con la (2.11) si determina *C* e con la (2.20) il valore di *Z* in *P₀*. La (2.23), scritta nel punto *P₀*, consente di ricavare *A*. Nelle nostre ipotesi *(0<φ<π)* è facile controllare che risulta *A>0*, perciò potremo porre:

$$A = e^{-\frac{\beta \bar{x}}{C}}, \qquad (2.24)$$

dando a $\bar{x}$ un opportuno valore (positivo, nullo o negativo). Risolvendo la (2.23) rispetto a *Z* otteniamo:

$$Z = \cosh[\frac{\beta(x-\bar{x})}{C}], \qquad (2.25)$$

da cui, tenendo presente la (2.20), segue:

$$z(x) = \frac{1}{\beta}\left\{C\cosh\left[\frac{\beta(x-\bar{x})}{C}\right] - n_{\min}\right\}. \qquad (2.26)$$

Dunque i raggi hanno, a meno di traslazioni verticali, la forma di coseni iperbolici o, come anche si dice, di catenarie. Naturalmente la (2.26) è valida nell'intervallo *0≤z≤z_max*.

**Bibliografia**

Balanis, C. A., (1989) *Advanced Engineering Electromagnetics*, Wiley.

Bertoni, H. L., (2000) *Radio Propagation for Modern Wireless Systems*, Prentice Hall.

Born, M., and Wolf, E., (1993) *Principles of Optics*, Cambridge University Press.

Felsen, L. and Marcuvitz, N., (1994) *Radiation and scattering of waves*, IEEE.

Kline, M. and Kay, I., (1965) *Electromagnetic Theory and Geometrical Optics*, Interscience.

Sommerfeld, A. J. W., (1954) *Optics*, Academic Press.